\documentclass[twocolumn,showpacs,preprintnumbers,amsmath,amssymb,prb]{revtex4}

\usepackage{graphicx}
\usepackage{bm}
\usepackage{subfigure}

\def\beq{\begin{eqnarray}}
\def\eeq{\end{eqnarray}}
\begin{document}
\draft

\title{Interference and Switching of Josephson current carried by nonlocal spin-entangled
electrons in a SQUID-like system with quantum dots}

\author{Zhi Wang and Xiao Hu}
\address{WPI Center for Materials Nanoarchitectonics, National Institute for
Materials Science, Tsukuba 305-0044, Japan}

\date{\today}

\begin{abstract}
Josephson current of spin-entangled electrons through the two
branches of a SQUID-like structure with two quantum dots exhibits a
magnetic-flux response different from the conventional Josephson
current. Due to their interference, the period of maximum Josephson
current changes from $h/2e$ to $h/e$, which can be used for
detecting the Cooper-pair splitting efficiency. The nonlocal spin
entanglement provides a quantum mechanical functionale for switching
on and off this novel Josephson current, and explicitly a switch is
formulated by including a pilot junction. It is shown that the
device can be used to measure the magnitude of split-tunneling
Josephson current.

\end{abstract}

\maketitle

\noindent {\it Introduction --} Nonlocal quantum entanglement
becomes the focus of many investigations in recent years. The
so-called Einstein-Podolsky-Rosen (EPR) pair\cite{epr} is the
hallmark of this phenomenon, which not only serves as the test case
of violation of the Bell inequality \cite{bell}, but also works as
the medium for quantum communication\cite{bennett} and quantum
computation\cite{nielsen}. Photons are most intensively
investigated, and their nonlocal entanglement has been successfully
demonstrated\cite{aspect,zhao}. It would be interesting to observe
this phenomenon in electron systems, where entanglement may arise in
either spin or spatial degrees of freedom. However, the generation
and detection of nonlocal entangled electrons in solid-state systems
is still a challenge, since electrons interact with the macroscopic
Fermi sea around them and it is hard to control a particular pair.

A possible approach to achieve nonlocal entanglement of an electron
pair is to take advantage of the intrinsically spin entangled Cooper
pairs in superconductivity. Based on this idea, three-terminal
devices consisting of an s-wave superconductor coupled to two leads
made of quantum dots (QDs), Luttinger liquid, and normal Fermi
liquid have been theoretically
proposed\cite{lossdot,lossluttinger,lossfermi}. It was shown that
the Coulomb blockade effect may split a Cooper pair and force
spin-entangled electrons tunnel into different leads. In a
SQUID-like structure these spin-entangled electrons generate a new
contribution to Josephson current\cite{lossdc}. The anticipated
Cooper pair splitting has then been explored experimentally in
mesoscopic systems\cite{beckmann04,russo05,zimansky,kleine,hofstetterNAT09,herrmannPRL10,wei09}.
The correlation between the resistances in the two leads due to the
crossed Andreev effect (CAR) was used to measure the efficiency of
Cooper pair splitting
\cite{andreev,byers95,falci01,lossdot,car2,car3}.

\begin{figure}[htb]
\begin{center}
\leavevmode
\includegraphics[clip=true,width=0.9\columnwidth]{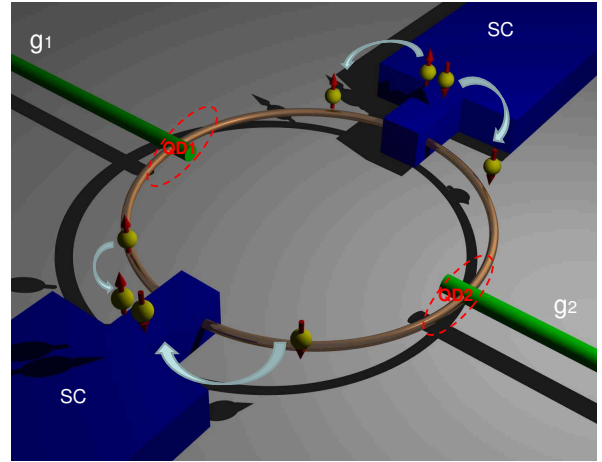}
\caption{Schematic setup of a Cooper pair splitter. Two
superconductors are connected by two leads with quantum dots (QDs)
embedded. The QDs are defined by voltage gate g$_1$ and g$_2$.}
\label{fig:setup}
\end{center}
\end{figure}

Generally speaking, however, the resistance correlation does not
provide a direct evidence for the nonlocal spin
entanglement\cite{wei09}, since processes without spin entanglement
may be involved\cite{falci01}. In the present work, we propose to
detect nonlocal spin entanglement based on the interference of
Josephson current, which is immune to any process involving
disentangled electrons. For this purpose, we adopt the system in
Fig.~1 which was first discussed by Ref.~10. We notice that the
novel Josephson current carried by spin-entangled electrons
tunneling through the two paths in Fig.~1 responds to the magnetic
flux differently from the conventional Josephson current for which a
Cooper pair tunnels through one of the two leads. Due to their
interference the maximum Josephson current exhibits a variation of
period $h/e$ responding to the magnetic flux, in contrast of the
conventional $h/2e$ known for SQUID, which can be used to detect the
splitting efficiency of Cooper pair. The nonlocal spin entanglement
is then shown to provide a quantum mechanical functionale for
switching on and off this split tunneling process. The function of a
switch based on a pilot Josephson junction is formulated explicitly.
It is shown that the device can be used to measure directly the
magnitude of the novel Josephson current.

\vspace{3mm} \noindent {\it Interference in presence of magnetic
flux --} It has been demonstrated experimentally
\cite{hofstetterNAT09} that QD can work as a notch to control the
Cooper pair splitting, since the Coulomb blockade effect suppresses
the tunneling of two electrons of a Cooper pair through the same QD.
The system can be described with a tunneling Hamiltonian
\cite{cohen} $ H=H_0+H_{\rm T}$, with $H_0=H_{\rm S}+H_{\rm QD}$,
where $H_{\rm S}=H_{\rm L}+H_{\rm R}$ is the BCS Hamiltonian of the
two superconductors,

\begin{eqnarray}
H_{\alpha}=\sum_{\sigma,{\bf k}}\xi_{\bf k}
c^{\dagger}_{\alpha,\sigma,{\bf k}}c_{\alpha,\sigma,{\bf
k}}+\sum_{{\bf k}} (\Delta_{\alpha} c^{\dagger}_{\alpha,{\bf
k},\uparrow}c^{\dagger}_{\alpha,-{\bf k},\downarrow}+h.c.)
\end{eqnarray}
with $\alpha=L, R$ for the left and right superconductor,
$\sigma=\uparrow, \downarrow$ for the electron spin; $H_{\rm
QD}=H_{\rm u}+H_{\rm d}$ is the Anderson-type Hamiltonian for the
QDs with one localized spin-degenerate energy level
\begin{eqnarray}
H_{\eta}=\epsilon_{\eta}\sum\limits_{\sigma}a_{\eta\sigma}^{\dagger}a_{\eta\sigma}+Un
_{\eta\uparrow}n_{\eta \downarrow}
\end{eqnarray}
with $\eta={u,d}$ for the up and down QDs. In the presence of
magnetic flux, the tunneling Hamiltonian is given by $H_{\rm
T}=H_{\rm Tu}+H_{\rm Td}$,
\begin{eqnarray}
H_{\rm Tu}=e^{-i\frac{\pi \Phi  }{4 \Phi_0}}T a_{{\rm
u}\sigma}^{\dagger}c_{{\rm R},\sigma}({\bf r}_{\rm u,R}) +e^{-i
\frac{\pi \Phi }{4 \Phi_0}}T c_{\rm L,\sigma}^{\dagger}({\bf r}_{\rm
u,L}) a_{{\rm u}\sigma}+{\rm h.c.}\nonumber
\end{eqnarray}
\begin{eqnarray}
H_{\rm Td}=e^{i \frac{\pi \Phi  }{4 \Phi_0}}T a_{{\rm
d}\sigma}^{\dagger}c_{\rm R,\sigma}({\bf r}_{\rm d,R}) +e^{i
\frac{\pi \Phi }{4 \Phi_0}}T c_{\rm L,\sigma}^{\dagger}({\bf r}_{\rm
d,L}) a_{{\rm d}\sigma}+{\rm h.c.},
\end{eqnarray}
with ${\bf r}_{\eta,\alpha}$ denoting the positions where QDs are
connected to superconductors, $T$ the tunneling matrix, $\Phi$ the
magnetic flux, and $\Phi_0=h/2e$ the flux quantuam.

The dc Josephson current can be evaluated using the Green function
technique, and we arrive at the following three contributions
$I_{\rm J}=I_{\rm u}+I_{\rm d}+I_{\rm ud}$:

\begin{eqnarray}
I_{\rm u}&&= \frac{4eT^2}{ \hbar} {\rm Re} \int
\frac{d\omega}{2\pi}n_{\rm F}(\omega)
 \\\ \times && [e^{i \frac{\pi \Phi }{2\Phi_0}} \Im^{ ret}_{\rm uu}(\omega)
\Im^{\dagger ret}_{\rm R}(\omega,{\bf 0}) -e^{ -i\frac{\pi \Phi
}{2\Phi_0}} \Im^{\dagger ret}_{\rm uu}(\omega) \Im^{
ret}_{\rm R}(\omega,{\bf 0})]\nonumber
\end{eqnarray}
\begin{eqnarray}
I_{\rm d}&&=\frac{4eT^2}{ \hbar} {\rm Re} \int
\frac{d\omega}{2\pi}n_{\rm F}(\omega)
\\\ \times && [ e^{ -i\frac{\pi \Phi }{2\Phi_0}} \Im^{ ret}_{\rm dd}(\omega)
\Im^{\dagger ret}_{\rm R}(\omega,{\bf 0})- e^{i \frac{\pi \Phi
}{2\Phi_0}} \Im^{\dagger ret}_{\rm dd}(\omega) \Im^{ret}_{\rm
R}(\omega,{\bf 0})]\nonumber
\end{eqnarray}
from Cooper pairs tunneling through the up and down QD respectively, which are
similar to the result for a single QD Josephson junction\cite{ishizaka}
except the Peierls factors, and
\begin{eqnarray}
I_{\rm ud} && =\frac{8eT^2}{\hbar} {\rm Re} \int
\frac{d\omega}{2\pi}n_{\rm F}(\omega)  \\\  \times && [\Im^{
ret}_{\rm ud}(\omega) \Im^{\dagger ret}_{\rm R}(\omega,\delta {\bf
r}_{\rm R})-\Im^{\dagger ret}_{\rm ud}(\omega) \Im^{ ret}_{\rm
R}(\omega,\delta {\bf r}_{\rm R})]\nonumber
\end{eqnarray}
from Cooper pairs which are split and tunnel coherently through
different QDs. Here $n_{\rm F}(\omega)$ is the Fermi distribution
function, $\delta {\bf r}_{\alpha}={\bf r}_{\rm u,\alpha}-{\bf
r}_{\rm d,\alpha}$ is the terminal distance, $\Im^{ret }_{\rm
R}(\omega,{\bf r}) \equiv \ll c_{\rm R,\uparrow}({\bf r});c_{\rm
R,\downarrow}({\bf 0})\gg$ is the retarded anomalous Green function
of the right superconductors and $\Im^{ ret }_{\eta \eta'}(\omega)
\equiv \ll a_{\eta,\uparrow};a_{\eta',\downarrow} \gg$ is the
retarded anomalous Green function of the QD. The superconductor is
assumed to be a macroscopic system and thus its anomalous Green
function is of the BCS form. The anomalous Matsubara Green function
of the QD can be calculated with a contour integral method,
\begin{figure}[b]
\begin{center}
\leavevmode
\includegraphics[clip=true,width=0.9\columnwidth]{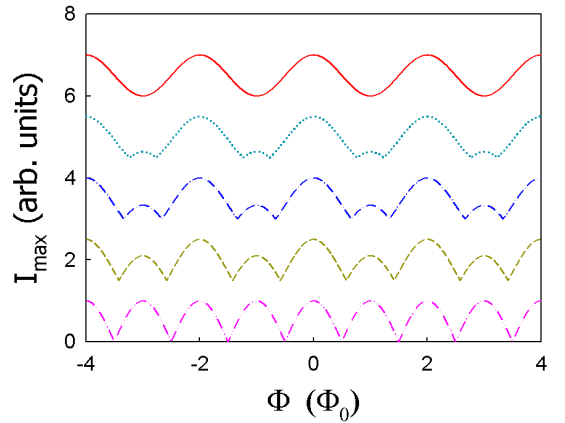}
\caption{(color online). Magnetic response of maximum dc Josephson
current. The dash-double dotted, dashed,
dash-dotted, dotted, solid line is for the case of
splitting efficiency $\gamma=$ 0, 0.25, 0.5, 0.75, 1 respectively.
Each curve was normalized and shifted for clarity.}
\end{center}
\end{figure}
\begin{eqnarray}
\Im^{\dagger}_{\eta \eta'}(i\omega) =\frac{1}{Z}  \int_C dz  e^{-
\beta z} {\rm Tr}[\frac{1}{z-H} a^{\dagger}_{\eta \uparrow}
\frac{1}{z+i\omega-H}a^{\dagger}_{\eta'\downarrow}]
\end{eqnarray}
where Z is the partition function and $\beta$ is the inverse
temperature. Here we consider the regime $\epsilon_{\rm
u}=\epsilon_{\rm d}=\epsilon>0$ where the ground state of the QD is
empty of electron, thus avoiding the complexity induced by the local
electrons on the QDs. Evaluating the Green function by perturbation
expansion in respect of the tunneling Hamiltonian, we arrive at the
lowest order result of the Josephson current at low temperatures

\begin{eqnarray}
I_{\rm J}=[I_1\sin(\phi_{\rm s}-\frac{\pi \Phi}{ \Phi_0})+
I_1\sin(\phi_{\rm s}+\frac{\pi \Phi}{ \Phi_0})+I_2\sin\phi_{\rm s}]
\end{eqnarray}
where $\phi_{\rm s}$ is the superconducting phase difference, and
\begin{eqnarray}
I_1= \frac{1 }{ \hbar}
 \sum_{{\bf k},{\bf p}}
\frac{2e T^4 |\Delta_{\rm L} \Delta_{\rm R}|}{E_{\bf k} E_{\bf
p}(E_{\bf k}+ \epsilon)(E_{\bf p}+\epsilon)}  [ \frac{1}{E_{\bf
k}+E_{\bf p}}
 +  \frac{1}{(2\epsilon+U)}
]
\end{eqnarray}
 with $E_k = \sqrt{\xi_k^2 + \Delta^2}$. It is diminished when the Coulomb interaction U and the
superconducting energy gap $\Delta$ are large\cite{ishizaka}, and
\begin{eqnarray}
I_2 &&=\frac{1}{\hbar}  \sum_{{\bf k},{\bf p}}  \frac{4e T^4
|\Delta_{\rm L} \Delta_{\rm R}| {\rm cos}({\bf k} \cdot \delta {\bf
r}_{\rm R} ) {\rm cos}({\bf p} \cdot \delta {\bf r}_{\rm L}
)}{E_{\bf k} E_{\bf p}(E_{\bf k}+\epsilon) (E_{\bf
p}+\epsilon)}\nonumber \\\
&& \times [\frac{1}{E_{\bf k}+E_{\bf p}}+\frac{1}{2\epsilon }]
\end{eqnarray}
which is controlled by the distances between the terminals of QD
channels\cite{lossdc}.  Here we note that the divergence
 of $I_2$ on $\epsilon$ is caused by
the perturbation treatment, which can be eliminated by proper renormalization\cite{ishizaka,rodero}.

There are three contributions in the Josephson current in Eq.~(8).
The first two terms come from Cooper pairs co-tunneling from the up
and down QD respectively. The interference between these two terms
results in a period of $\Phi_0$ in the magnetic response of maximum
Josephson current, identical to that of a conventional SQUID. The
third term is unique for the present system, in which the two
electrons of a Cooper pair are split and tunnel through different
QDs, and the Peierls factors cancel each other exactly. As the
result, the interference pattern of maximum dc Josephson current
exhibits a period of $2\Phi_0$. It is clear that the behavior of the
total Josephson current is significantly modulated by the process of
Cooper pair splitting. In order to see this influence more clearly,
we define the splitting efficiency $ \gamma \equiv {I_2}/{2{I_1}}$.
In the regime
of strong Coulomb interaction and single particle resonate tunneling
with $\epsilon\ll\Delta\ll U$, one has
 $\gamma \approx \frac{\pi \Delta}{(4-\pi) \epsilon } e^{- 2{\delta
{\bf r}}/{\pi \xi}}\sin^2(k_{\rm F} \delta {\bf r})/
(k_{\rm F} \delta {\bf r})^2$ , where $\delta {\bf r}=\delta {\bf
r}_{\rm L}=\delta {\bf r}_{\rm R}$, where $\delta {\bf r}=\delta {\bf
r}_{\rm L}=\delta {\bf r}_{\rm R}$, $\xi$ is the coherence length
and $k_F$ is the Fermi wave vector. Theoretically, the factor $\sin^2(k_{\rm F} \delta {\bf r})/
(k_{\rm F} \delta {\bf r})^2$ sets the common condition for observing
the present quantity and the cross Andreev reflection \cite{lossdot,falci01}. The
variation of maximum Josephson current with the splitting efficiency
is illustrated in Fig.~2. It is clear that with increasing splitting
efficiency from zero to unity, the pattern evolves gradually, and
the period changes from $\Phi_0$ to $2 \Phi_0$. It is noticed that
the splitting efficiency can be larger than unity, associated with a
similar curve to $\gamma=1$, except that the minima do not reach
zero.

\vspace{3mm} \noindent {\it Switching of the novel Josephson current
--}  In order to contribute to the split-tunneling Josephson
current, two single electrons tunneling through the two paths should
be in a spin singlet state. This strong spin entanglement can be
used for switching on and off this novel Josephson current. Let us
set $-U/2<<-\Delta<\epsilon<0$ such that there is a localized
electron on each QD \cite{lossdc}. For the split-tunneling Josephson
current, it is the "on"/"off" state when the spins of the two
localized electrons are in a spin singlet/triplets state. The
situation can be described by \cite{lossdc}
\begin{figure}[t]
\begin{center}
\leavevmode
\includegraphics[clip=true,width=0.9\columnwidth]{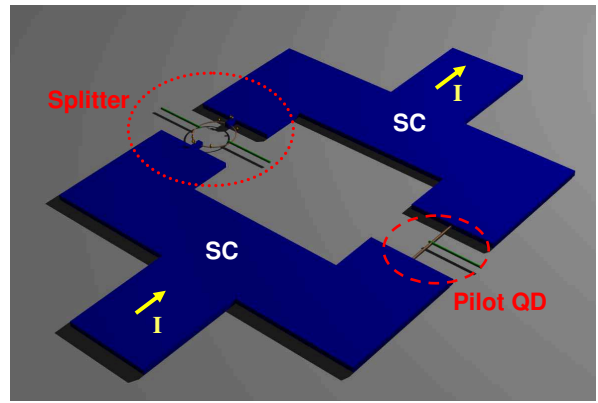}
\caption{(color online). Schematic setup of a switch of the novel
Josephson current. The Cooper pair splitter displayed in Fig.~1 is
connected in parallel with a pilot Josephson junction.}
\end{center}
\end{figure}
\begin{equation}
I=I_2 (\frac{1}{4} -{\bf S}_{\rm u} \cdot {\bf S}_{\rm
d})\sin\phi_{\rm s}.
\end{equation}

Here we propose a way to control the two localized spins by
introducing an additional pilot QD-junction as shown schematically
in Fig.~3. The critical current $J$ of the pilot junction varies
continuously when the gate voltage on the QD is tuned\cite{doh,dam}.
The total Josephson current is then given by

\begin{eqnarray}
I_{\rm J}=[2I_1 + I_2 (\frac{1}{4} -{\bf S}_{\rm u} \cdot {\bf
S}_{\rm d})+J]\sin\phi_{\rm s}
\end{eqnarray}
with the co-tunneling current\cite{ishizaka}

\begin{eqnarray}
I_1 = -\frac{1}{\hbar}
 \sum_{{\bf k},{\bf p}} \frac{2eT^4|\Delta_{\rm L} \Delta_{\rm R}|}{E_{\bf k} E_{\bf p}(E_{\bf k}+E_{\bf p})(E_{\bf k}-\epsilon)(E_{\bf p}-\epsilon)}
\end{eqnarray}
and the split-tunneling current\cite{lossdc}
\begin{eqnarray}
I_2 &&=\frac{1}{\hbar}  \sum_{{\bf k},{\bf p}}  \frac{4e T^4
|\Delta_{\rm L} \Delta_{\rm R}| {\rm cos}({\bf k} \cdot \delta {\bf
r}_{\rm R} ) {\rm cos}({\bf p} \cdot \delta {\bf r}_{\rm L}
)}{E_{\bf k} E_{\bf p}(E_{\bf k}-\epsilon) (E_{\bf
p}-\epsilon)}\nonumber \\\
&& \times [\frac{1}{E_{\bf k}+E_{\bf p}}+\frac{1}{2|\epsilon| }]
\end{eqnarray}
where only the zero order terms in U are included for simplicity,
since U is the largest energy scale. Both $I_1$ and $I_2$ are
similar to the previous section, except that $I_1$ is negative,
which represents the $\pi$ junction nature of each QD for
co-tunneling current. For most cases, $-I_1 > I_2$.

The critical current of the total system can be straightforwardly
evaluated for the two spin configurations, which is plotted as
function of the critical current $J$ of the pilot junction in
Fig.~4. For $J<-2I_1-I_2/2$, the critical current of the spin
triplets state is larger than that of the singlet state, and vice
versa.

When a current is injected into the system through one of the
superconductors, the superconductivity phase difference will be
adjusted in order to pass this current without dissipation. If the
current is small, the localized spins can take either singlet or
triplet state. For current in between the two critical currents, the
system will adjust the localized spins to achieve the larger
critical current in order to reduce dissipation\cite{lossdc}. It is easy to
figure out that for $J<-2I_1-I_2/2$ the spin
triplets, thus "off", state is realized, and for $J>-2I_1-I_2/2$ the
spin singlet, thus "on" state, is realized, as depicted in Fig.~4.

 It is noticed that
there is a gap $I_g$ in the maximal Josephson current of the total
system when the critical current of the pilot junction
is swept, due to the split-tunneling Josephson current. By measuring
the gap current we can evaluate the magnitude of the split-tunneling
Josephson current since $I_2=2I_{\rm g}$. At the switching point,
the Josephson current carried by co-tunneling processes is
suppressed significantly down to $-I_2/2$, and thus the splitting
efficiency of Cooper pair is enhanced largely.

\begin{figure}[t]
\begin{center}
\leavevmode
\includegraphics[clip=true,width=1.\columnwidth]{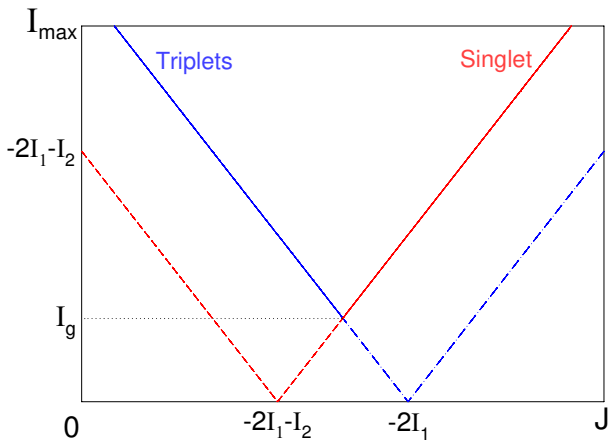}
\caption{(color online). Maximum Josephson current of the total
system in response to critical current of the pilot junction $J$.
The dashed/dash-dotted line is for the spin triplets/singlet state,
and the solid lines are for the states realized. }
\end{center}
\end{figure}

Before ending this section, we notice that the pilot QD can be replaced
by a conventional SQUID, for which the critical current can be controlled
by a magnetic field.

\vspace{3mm} \noindent {\it Summary -- } In summary, we analyze the
tunneling processes in a SQUID-like device with a quantum dot
embedded in each junction, where electrons with nonlocal spin
entanglement are generated and tunnel through different junctions,
which contribute to the Josephson current. In presence of a magnetic
flux, they carry a zero Peierls phase due to the cancelation between
the two tunneling paths, in contrast to the tunneling of Cooper
pairs through one of the two junction. As the splitting efficiency
increases, the period of the magnetic-flux response of maximum
Josephson current changes from $\Phi_0=h/2e$ to $2\Phi_0$. The
nonlocal spin entanglement is then used to switch the split
tunneling process. The "on" ("off") state of the switch corresponds
to the spin singlet (triplet) of the two localized electrons in the
quantum dots of the Cooper-pair splitter, which can be controlled by
a pilot junction. It is shown that the device can be used to measure
directly the magnitude of the Josephson current carried by single
electrons with nonlocal spin entanglement.

\vspace{3mm} \noindent {\it Acknowledgements -- } The authors thank
H. Takayanagi and S.-Z. Lin for helpful discussions. This work was
supported by WPI Initiative on Materials Nanoarchitectonics, MEXT, and
partially by Grants-in-Aid for  Scientific Research (No.22540377), JSPS,
and CREST, JST, Japan.

\end{document}